# Graphene Side Gate Engineering

Ching-Tzu Chen, Tony Low, *Member, IEEE*, Hsin-Ying Chiu, and Wenjuan Zhu, *Member, IEEE*

*Abstract*— Various mesoscopic devices exploit electrostatic side gates for their operation. In this paper, we investigate how voltage-biasing of graphene side gates modulates the electrical transport characteristics of graphene channel. We explore myriads of typical side gated devices such as symmetric dual side gates and asymmetric single side gate biasing, in monolayer and bilayer graphene. The side gates modulate the electrostatic doping in the graphene channel whose effect is reflected in transport measurement. This modulation efficiency is systematically characterized for all our devices and agrees well with the modeling presented.

## I. Introduction

GRAPHENE side gates have been applied extensively in mesoscopic graphene quantum devices such as single-electron transistors[1] and quantum dots,[2] as the plunger gates to vary the electron number on the dot and as the tunable barrier to control the electron tunneling rate. Furthermore, in graphene field-effect devices,[3,4] the side gates could offer a better alternative to top-gating scheme as a means to modulate the channel doping as it avoids dielectric breakdown and hysteresis caused by top-gate dielectrics.[5,6] Design of side gated graphene devices hinges crucially upon our understanding of how the side-gate-induced fringing fields impact the electrical transport. Investigations of this issue are few[7,8] to date. In this letter, we conduct a systematic study of graphene devices with different side gating configurations. Through the measured transport characteristics, we are able to quantify the modulation efficiency of the side gates. Numerical simulations are employed in understanding and quantifying the side gate modulation efficiency seen in experiments.

## II. Device Design and Fabrication

To begin, we model the electrostatics of a prototypical graphene side-gate device as shown in figure 1(a). The channel length $L$ of the device is ~2.0μm, channel width $W$ ~ 400nm, and channel-edge to side-gate distance $d$ ~ 200nm. Since $L$ is considerably larger than $W$ and $d$, we ignore the field component along the length direction and simulate the electrostatics for a cross-section of our device as indicated. Fig. 1(b) depicts the 2D potential profile at the bias point ($V_{bg}$, $V_{sg}$) = (0V, 30V) calculated using finite-element method, with the channel grounded. From the potential distribution, we then derive the induced charge-density along the graphene channel for each ($V_{bg}$, $V_{sg}$) bias. Fig. 1(c) demonstrates the impact of side-gate bias on channel charge distribution. As we can see, applying side-gate voltages modulates the carrier density in substantial part of the channel. This will subsequently affect

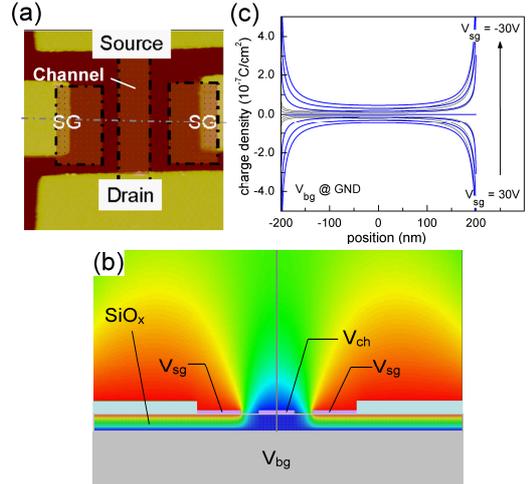

Fig. 1. (a) AFM micrograph of a prototypical side-gate device. The device channel length $L$ ~ 2.0μm, channel width $W$ ~ 400nm, and channel-edge to side-gate distance $d$ ~ 200nm. (b) Simulated 2D electrostatic potential profile along the dash-dot line cut (in gray) of the device in (a) at ($V_{bg}$, $V_{sg}$) = (0V, 30V). The x-axis points along the direction of the line cut, and y-axis points out of the graphene plane. The graphene channel and side-gates are highlighted in light magenta, and the metal contacts (~40nm thick) to the side-gates in light cyan. (c) Charge density profile along the graphene channel in the symmetric (blue) side-gating scheme and the asymmetric (black) scheme where the side gate on the left is grounded. Charge density is derived from the difference between the out-of-plane component of the electric displacement field above and that below the channel as a function of increasing $V_{sg}$ in intervals of 10V, with $V_{bg}$ fixed at ground potential.

the electrical transport behavior, to be discussed below.

We proceed with the graphene side-gate device fabrication by standard e-beam lithography (EBL) of exfoliated monolayer and bilayer graphene. Graphene flakes are deposited on highly doped Si substrate capped by ~90nm of thermal oxide. The source and drain contacts are first patterned using EBL, e-beam deposition of Ti/Pd/Au (1nm/20nm/20nm), and lift-off. The graphene channels and side gates are then defined using oxygen reactive-ion etching in one step. For bilayer devices with an additional top gate, silicon nitride is chosen as the dielectric because of its high yield and high breakdown field. Plasma-enhanced CVD is used to deposit 30nm of silicon nitride at 400°C,[9] and the final EBL and metallization steps form the top-gate electrode.

## III. Results and Discussion

### A. Transport Measurements

Transport measurements are carried out at ~5K with a low source-drain bias ($V_{ds}$=0.1V). Fig. 2 shows the representative datasets under various gating schemes. In each device, we record the resistance ($R$) v.s. back-gate voltage ($V_{bg}$) curves in constant intervals of the side-gate voltage ($V_{sg}$). As expected, we see a pronounced change in the device $R$-$V_{bg}$ characteristics with applied $V_{sg}$, the most salient feature of which is the $V_{sg}$-dependent shift in the back-gate voltage corresponding to maximum channel resistance ($V_{bg}^*$). To quantify the side-gate modulation on the device channel, we map out $V_{bg}^*$ for each $V_{sg}$ as illustrated in the inset of Fig. 2(b). $V_{bg}^*$ depends roughly linearly on $V_{sg}$. Thus we define parameter $\eta = |\Delta V_{bg}^*/\Delta V_{sg}|$ as the measure of the side-gate modulation efficiency.

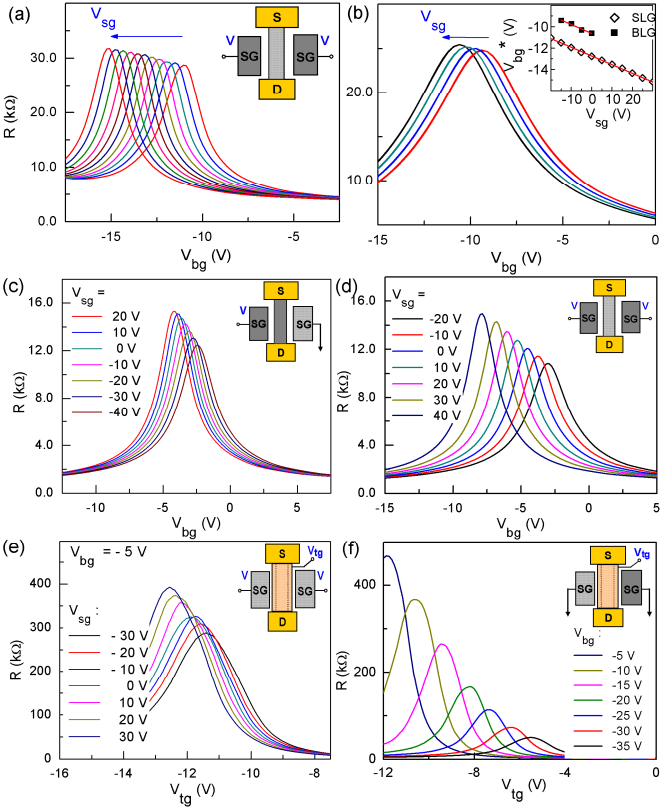

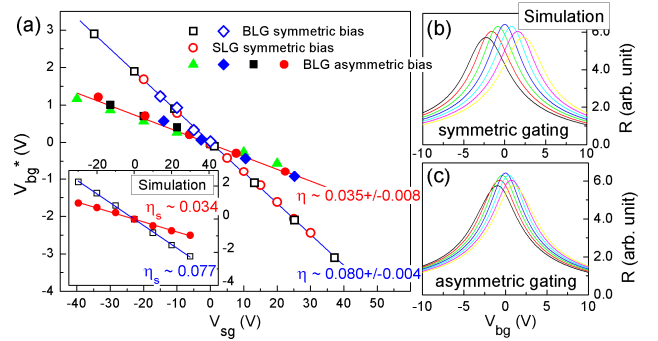

Fig. 2. Side-gate dependence of resistance ($R$) vs. back-gate voltage ($V_{bg}$) transport characteristics in monolayer and bilayer devices. (a) $R$-$V_{bg}$ plot of a symmetrically gated monolayer device (SLG). Graphene side-gate voltage $V_{sg}$ varies from -20V to 30V in steps of 5V. (Arrow indicates the direction of increasing $V_{sg}$.) Inset: schematics of the side-gate biasing scheme. (b) $R$-$V_{bg}$ plot of a symmetrically gated bilayer device (BLG) with nominally the same device dimensions. $V_{sg}$ varies from -15V to 0V in steps of 5V. (Arrow indicates increasing $V_{sg}$.) Inset: $V_{bg}^*$- $V_{sg}$ plot, depicting the modulation efficiency of graphene side gate on $V_{bg}^*$, the maximum channel resistance point. (c) $R$-$V_{bg}$ plot of a side-gated bilayer device under asymmetric (single-sided) $V_{sg}$ gating; (d) $R$-$V_{bg}$ curves of a side-gated bilayer device under symmetric (double-sided) $V_{sg}$ gating; (e) $R$-$V_{tg}$ curves of a top-gated bilayer graphene device under symmetric (double-sided) graphene $V_{sg}$ gating with a fixed $V_{bg}$; (f) $R$-$V_{tg}$ curves of a top-gated bilayer device with increasing $V_{bg}$, keeping $V_{sg}$ grounded; the maximum resistance modulation is the manifestation of gap opening under a vertical electric field.

Fig. 3. Side-gate efficiency measured by the change in the maximum resistance point $V_{bg}^*$ with applied $V_{sg}$. (a) Main panel: Experimental $V_{bg}^*$-$V_{sg}$ data and linear fits, in which the modulation efficiency of the symmetric bias configuration is found to be $\eta \sim 0.08\pm0.004$ and that of the asymmetric single-gated bias configuration $\eta \sim 0.035\pm0.008$. Inset: Simulated $V_{bg}^*$-$V_{sg}$ data (with effective $SiO_x$ thickness $\sim$ 100nm) and linear fits, in which the $\eta$ of the symmetric bias configuration is found to be $\eta \sim 0.077$ and that of the asymmetric bias configuration $\eta \sim 0.034$. (b), (c) Simulated $R$-$V_{bg}$ transport characteristics with $V_{sg}$ varying from -30V to 30V in steps of 10V in the (b) symmetric and (c) asymmetric biasing scheme.

We first compare the response of monolayer (Fig. 2(a)) with that of bilayer (Fig. 2(b)) graphene channels to the voltage $V_{sg}$ applied symmetrically to the graphene side gates. In what we call the symmetric biasing configuration, both side gates are held at equal potential. (See inset of Fig. 2(a).) Figs. 2(a) and 2(b) address whether the difference in electronic band structure of monolayer and bilayer graphene causes observable differences in their response to side gate biasing in terms of the modulation efficiency $\eta$. We find that both in monolayer and bilayer devices, the maximum resistance point $V_{bg}^*$ shifts linearly toward the negative bias owing to the induced channel charges. More importantly, the side-gate modulation efficiency is comparable in the two types of devices ($\eta \sim 0.080$). Electrostatically, the carrier density profile in response to the side gating is the same in monolayer and bilayer channels. Their similar modulation efficiency $\eta$ suggests that $\eta$ has more to do with simple electrostatics (i.e. carrier density) than the actual electronic structure, as will be elucidated in III.*B*.

Having established that the side-gate modulation on monolayer and bilayer channels is comparable, we then study the modulation efficiency $\eta$ of various gating schemes. Figs. 2(c) and 2(d) compare the transport characteristics of bilayer channel under asymmetric and symmetric side-gate biasing. (The schematics in the top-right corner of each panel depict the biasing configurations.) In both cases, $V_{bg}^*$ varies linearly with $V_{sg}$, but the gating efficiency $\eta$ is significantly reduced in the asymmetric single-gate biasing configuration regardless of the polarity of the side-gate voltage. This decrease in modulation efficiency originates from a much smaller induced carrier density when one of the side gates is grounded (c.f. Fig. 1(c)). In Fig. 3, we summarize the $V_{bg}^*$-$V_{sg}$ relation mapped out from Fig. 2(a)-(d) and similar devices. The datasets are vertically offset so that they collapse onto each other for comparison purposes. We extract $\eta$ for different channel types and side-gate schemes using linear fitting. We find that the symmetric dual-gate configuration ($\eta \sim 0.080 \pm 0.004$) is more than twice as efficient as the asymmetric single-gate configuration ($\eta \sim 0.035 \pm 0.008$).

Lastly, we measure the side-gate modulation effect on the top-gated bilayer device. (See Figs. 2(e) and 2(f).) The side-gate efficiency $\eta$ is derived as follows. First, we obtain $|\Delta V_{tg}*/\Delta V_{bg}| \sim 0.21$ (where $V_{tg}*$ is the top-gate voltage of the maximum resistance point) from the $R$-$V_{tg}$ curves at $V_{sg} = 0V$ in Fig. 2(f), in which the substantial enhancement of $R$ is the manifestation of a perpendicular-E-field-induced gap. We then fix $V_{bg}$ (at -5V) and measure $R$-$V_{tg}$ in constant intervals of $V_{sg}$ (Fig. 2(e)) to obtain $|\Delta V_{sg}/\Delta V_{tg}*| \sim 53$. We thus find that, in the top-gated device, the side-gate biasing $\eta \sim 0.089$ proves equally effective in modulating channel transport.

*B. Experimental vs. Numerical Results*

Previously, we show by electrostatics simulation that side-gate biasing can induce doping variations within the graphene channel. Experimentally, this effect is reflected in the transport behavior of our devices. Here, we employ the well-known Landauer transport model to account for the measured device transport characteristics. Landauer's formula in the quasi-diffusive regime is given by $G = R^{-1} = (2e^2/h) \cdot M \cdot L_0/(L+L_0)$, where $L_0$ is the electron mean-free path, $M \sim k_f \cdot W/\pi = \sqrt{(n/\pi)} \cdot W$ is the number of transport modes and $n$ is the carrier density. From the electrostatic simulation in Sec. II, we obtain a spatially dependent charge density, i.e. $n(x)$. An effective mode number can then be computed i.e. $M = <M(x)>$. The quantity $L_0$, which in general is doping dependent, can be deduced from typical empirical $R$-$V_{bg}$ data at $V_{sg} = 0$. The extracted $L_0$ of our devices approaches 100nm in the linear-conductance limit away from the charge-neutrality point, consistent with graphene transport under the influence of screened charge-impurity scattering. With the device $L_0$ data, the channel conductance can then be calculated from the Landauer formula, and the $V_{sg}$-modulated $R$-$V_{bg}$ curves for the symmetric and asymmetric gating configurations are shown in Figs. 3(b) and 3(c). Note that the relation between $M$ and $n$ stated above applies both for monolayer and bilayer graphene. Despite the simplicity of this phenomenological model, it captures the experimental trend and shows similar degree of modulation as the experimental curves. Again, we plot the maximum resistance point $V_{bg}*$ for every $V_{sg}$ of the simulated data in the inset of Fig. 3(a) and compare them with the experimental $V_{bg}*$-$V_{sg}$ data in the main panel of Fig. 3(a). Using linear fitting, we see that the side-gate efficiency of the symmetric side-gate biasing ($\eta = 0.077$) and that of the asymmetric biasing ($\eta = 0.034$) extracted from the simulation corroborates the experimentally measured efficiency of 0.080±0.004 and 0.035±0.008. Improvements in $\eta$ could result from employing a higher dielectric medium and engineering of the side gates proximity. For instance, with a fixed channel width $W = 400$nm, reducing the channel-edge to side-gate distance from $d = 200$nm to 50nm increases the modulation efficiency by ~140% to 0.183 (Fig. 4(a)). Furthermore, even scaling down $W$ at a constant $d$ helps improving the side-gating efficiency. Given $d = 200$nm, $\eta$ increases to 0.154 as we trim the channel width down to $W = 50$nm (Fig. 4(b)), showing the effectiveness of side-gating

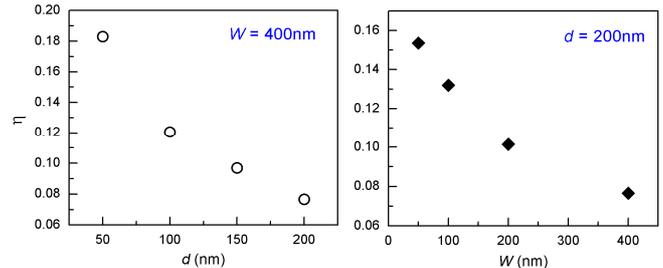

Fig. 4. Simulated improvement in side-gate modulation efficiency $\eta$ with scaled down device parameters in the symmetric-biasing configuration. (a) $\eta$ as a function of device channel-edge to side-gate distance $d$, at a constant channel width ($W = 400$nm). (b) $\eta$ as a function of device channel width $W$, at a constant channel-edge to side-gate distance ($d = 200$nm).

in the narrow-channel limit.

IV. CONCLUSION

We demonstrate, in various gating schemes, how graphene side gates can impact device electrostatics and ultimately modulate the device transport characteristics. We quantify the strength of modulation in terms of the side-gate efficiency parameter $\eta$ extracted from the transport data, and find that the measurement results agree very well with the simulated efficiency. Therefore, the experimental and modeling studies presented here provide a basis for designing general graphene planar multi-gate schemes for mesoscopic quantum devices and field-effect transistors.

ACKNOWLEDGMENT

We are grateful to D. Newns for illuminating discussions. We thank S.J. Han, D. Neumayer, B. Ek, J. Buccagniano, and G. Wright for expert help in device fabrication.